\documentclass{emulateapj}
\usepackage{times}

\newcommand{\beq}{\begin{equation}}
\newcommand{\eeq}{\end{equation}}

\def\farcd{\hbox{$.\mkern-4mu^\circ$}}
\def\farcm{\hbox{$.\mkern-4mu^\prime$}}

\def\earth{\hbox{$\oplus$}}
\def\arcmin{\hbox{$^\prime$}}
\def\arcsec{\hbox{$^{\prime\prime}$}}
\def\solar{\mbox{$_{\normalsize\odot}$}}

\def\deg{\hbox{$^\circ$}}

\newcommand{\AmS}{{\protect\the\textfont2
  A\kern-.1667em\lower.5ex\hbox{M}\kern-.125emS}}
\newcommand{\lsim}{\ \raise
-2.truept\hbox{\rlap{\hbox{$\sim$}}\raise5.truept\hbox{$<$}\ }}
\newcommand{\gsim}{\ \raise
-2.truept\hbox{\rlap{\hbox{$\sim$}}\raise5.truept\hbox{$>$}\ }}
\newcommand{\simsim}{\ \raise
-2.truept\hbox{\rlap{\hbox{$\sim$}}\raise5.truept\hbox{$\sim$}\ }}

\slugcomment{Accepted for Publication in the Astrophysical Journal, 6
November 2008}

\righthead{A New Diagnostic Method for Stellar Stratification}
\lefthead{Gouliermis, de~Grijs, \& Xin}

\shorttitle{A New Diagnostic Method for Stellar Stratification}
\shortauthors{Gouliermis, de~Grijs, \& Xin}

\begin{document}
 
\title{A New Diagnostic Method for Assessment of Stellar Stratification
in Star Clusters}

\author{Dimitrios A. Gouliermis}
\affil{Max-Planck-Institut f\"{u}r Astronomie, K\"{o}nigstuhl 17,
D-69117 Heidelberg, Germany\\ dgoulier@mpia.de}

\and

\and
\author{Richard de Grijs and Yu Xin}   
\affil{Department of Physics \& Astronomy, The University of Sheffield,
Hicks Building, Hounsfield Road, Sheffield S3 7RH, UK\\
R.deGrijs@sheffield.ac.uk; xinyu@bao.ac.cn}


\begin{abstract} 
We propose a new method for the characterization of stellar
stratification in stellar systems. The method uses the mean-square
radius (also called the {\sl Spitzer radius}) of the system as a
diagnostic tool. An estimate of the observable counterpart of this
radius for stars of different magnitude ranges is used as the {\em
effective radius} of each stellar species in a star cluster. We explore
the dependence of these radii on magnitude as a possible indication of
stellar stratification. This method is the first of its kind to use a
dynamically stable radius, and though seemingly trivial it has never
been applied before.  We test the proposed method using model star
clusters, which are constructed to be segregated on the basis of a Monte
Carlo technique, and on {\sl Hubble Space Telescope} observations of
mass-segregated star clusters in order to explore the limitations of the
method in relation to actual data. We conclude that the method performs
efficiently in the detection of stellar stratification and its results
do not depend on the data, provided that incompleteness has been
accurately measured and the contamination by the field population has
been thoroughly removed. Our diagnosis method is also independent of any
model or theoretical prediction, in contrast to the `classical' methods
used so far for the detection of mass segregation.
\end{abstract}


\section{Introduction}

{\sl Mass segregation} or {\sl stellar stratification} is the
phenomenon, according to which the massive stars in a star cluster are
mostly concentrated toward its center. Due to dynamical evolution
massive stars move inwards to the center of the cluster through two-body
interactions. Due to the same process less massive stars may move
outwards, so that on average the system `evaporates' (e.g., Lightman \&
Shapiro 1978; Meylan \& Heggie 1997). This is more commonly known as
{\sl dynamical} mass segregation. However, mass segregation is also
being observed in dynamically young stellar systems, supposedly due to
the formation of massive stars closer to the center of the newly-born
system (e.g., Murray \& Lin 1996; Bonnell \& Davies 1998). This type of
mass segregation is usually referred to as {\sl primordial}.

Both types of stellar stratification affect the physical properties of
the cluster in a manner which allows us to develop diagnostic tools
for its detection and quantification. Such tools involve the
identification of changes in the surface density profiles and the mass
functions of the clusters, or of the core radii of stars in different
mass groups. Specifically, stellar stratification in a star cluster
can be detected from the surface density profiles of stars in
different brightness ranges. If $f$ is the stellar surface density
(for every brightness range) within a distance $R$ from the center of
the cluster, then $f(R)$ can be used as an indication of
stratification at the magnitude limit, where it changes significantly
(e.g., Scaria \& Bappu 1981; Sagar et al. 1988; Subramaniam et al.
1993; Kontizas et al.  1998). The density profiles can be approximated
by $\log{f} \propto \gamma \log{R}$ (Elson et al. 1987) and the slope,
$\gamma$, can be used to quantify any significant change in
$f(R)$. Although this approximation was well designed, in practice
peculiarities of the density profiles toward the center of clusters
that are not completely spherical tend to produce large uncertainties
in $\gamma$, which hide any potential mass segregation (see, e.g.,
Gouliermis et al. 2004).

Another well-applied method is the construction of the luminosity or
mass function (LF, MF) of the cluster and the investigation of its
radial dependence (e.g., Pandey et al. 1992; King et al. 1995; Fischer
et al. 1998; Kontizas et al. 1998, de Grijs et al. 2002a,b). If the
cluster is indeed segregated, then the appearance of more massive
stars toward its center will result in shallower LFs and MFs toward
smaller radii (e.g., de Grijs et al. 2002a). However, this method has
several constraints imposed by the difficulties in the construction of
an accurate MF based on the use of theoretical mass-luminosity (ML)
relations. The use of different ML relations may result in significant
differences in the MF slopes obtained, independently of whether any
mass segregation is actually present (de Grijs et al. 2002b). In
addition, the investigation of any radial dependence of the LF and MF
slopes is not as straightforward as it seems, since its detection is
rather sensitive to the radial distance range being considered, i.e.,
where the slope is to be estimated (Gouliermis et al. 2004).

A rather interesting method is based on the use of the core radius of
the cluster as estimated for stars in specific magnitude (or mass)
ranges (e.g., Brandl et al. 1996; de Grijs et al. 2002b). Core radii
can be derived from the corresponding stellar number-density profiles
by fitting to the latter the models of Elson et al. (1987) and by
approximating them to the canonical King models for Galactic globular
clusters (King 1966). However, this method seems to be model-dependent
to a degree, which makes its application difficult. Another proposed
method is to measure the mean stellar mass within a given radius
(e.g., Bonnell \& Davies 1998; Hillenbrand \& Hartmann 1998), but this
diagnostic requires accurate knowledge of the stellar masses, which is
not always feasible (de Grijs et al. 2002b; Gouliermis et al. 2004).

For the methods discussed above, detecting stellar stratification
requires observations at high resolution, so that detailed radial
density distributions for the fainter magnitude ranges and complete
MFs in several annuli around the center of the cluster can be
constructed accurately. These methods can provide useful information
about the segregated masses as well as on the radius up to which the
cluster is shown to be segregated. However, it seems that all of these
methods are very sensitive to the quality of the observed data, as
well as to model assumptions. Moreover, all of the methods for the
detection of stellar stratification used thus far are closely
associated to the fitting of predefined functions to the data and
searching for parametric differences in these functions. One such
method, for example, involves the study of the stellar projected
density distribution of stars of different magnitudes; here, the
differences in the exponents of the slopes fitted are seen as proof of
mass segregation. A second method is based on the inherently imperfect
derivation of radius-dependent MFs, while yet another makes use of the
core radius of the system for different stellar groups, relying on
apparent structures, but it is extensively model-dependent. The
problem that all these studies suffer from is that {\sl it is very
difficult to compare their results among each other, thus leading to
extensive discussions about the nature and even the reality of stellar
stratification}.

In this paper we present a robust method for the detection of stellar
stratification. The simple application of this method can yield direct
information about the spatial extent (radial distribution) of stellar
groups in different magnitude (mass) ranges. It is based on the notion
of a dynamically stable radius of a star cluster, the so-called
{\sl `Spitzer radius,'} and leads to an observed {\sl `effective
radius.'} The diagnostic method for stellar stratification is
described in \S~2. The diagnostic process is applied to and tested on
a set of simulated spherical star clusters in \S~3, which are
constructed on the basis of Monte Carlo simulations. We thus validate
the use of the diagnostic and the unavoidable constraints. We then
apply the method to real data of two intermediate-age clusters in the
Magellanic Clouds observed with the {\sl Advanced Camera for Surveys}
(ACS) on-board the {\sl Hubble Space Telescope (HST)}, and we report on
our results in \S~4. A comparison of the results of our method with
those of the classical method of the MF dependence for the test
clusters is also presented in \S~4. Conclusions and a discussion on
the use of this diagnostic tool are summarized in \S~5.


\section{Diagnostic for Spatial Stellar Stratification}\label{s:method}

\subsection{The Spitzer Radius}\label{s:spitztheory}

The radius commonly known as the {\sl `Spitzer Radius'} of a star
cluster was introduced by L. Spitzer Jr. and his collaborators, who
used it as a distance indicator for different stellar subclasses in a
cluster (see, e.g., Spitzer 1958). Since gravity operates proportional
to the inverse square of the distance between a dynamical center and a
subject mass, if the latter represents a group of masses distributed
around the center (e.g., stars in a cluster), then the square distance
of this group can be replaced by the mean-square distance of its
members. Specifically, the gravitational force in a spherical stellar
system at a distance $r$ from its center, is (e.g., Spitzer 1958,
1969)
\beq \frac{dV}{dr} = \frac{GM_{r}}{r^{2}} ,
\label{app-a1} \eeq 
where $V$ is the potential and $M_{r}$ the mass contained within a
radius $r$. King (1965) showed that considering only one single type
of stars with mass $m$ and if $S(x)dx$ is the number of stars in
a strip of width $dx$, then assuming a Plummer (1911) stellar
distribution, the potential is
\beq V(r)=-\frac{2Gm}{r}\int_{0}^{r}S(x)dx{\rm .} \label{app-a2}
\eeq 
Furthermore, Spitzer, investigating energy equipartition between two
different mass groups, notes that the term $r^{2}$ in Eq.
(\ref{app-a1}) may be expressed as the mean value of the square
distance of all stars in a single mass group, if they were the only
cluster members. Consequently, the square root of this value gives $r$
in Eq. (\ref{app-a2}). This radius, later called the `Spitzer radius,'
corresponds to the distance up to which the stars of this specific
mass group affect the gravitational field of the cluster as a whole.

Therefore, the Spitzer radius of a star cluster is defined as the
mean-square distance of the stars from its center, \beq r_{\rm
Sp}=\sqrt{\frac{\displaystyle\sum^{N}_{i=1} {r_{i}}^{2}}{N}} ,
\label{eq-01} \eeq where $r_{i}$ is the radial distance of the $i^{\rm
th}$ star and $N$ is the total number of stars. For a Maxwellian
distribution of a group of masses in a parabolic potential well, which
represents that of a globular cluster very well, the half-mass radius
of the system is almost equal to ($\sim$90\% of) the Spitzer radius
(Spitzer 1969). Consequently, considering the limitations in the
measurement of the half-mass radius of a cluster versus the direct
measurement of the Spitzer radius, the latter can be considered quite
important in relation to the dynamical status of the cluster, being
used as its characteristic radius.

As far as stellar stratification is concerned, in a star cluster
displaying mass segregation one should be able to observe the more
massive stars concentrated toward the center of the system by plotting
the position coordinates of the stars according to the their
magnitudes. The Spitzer radius, being a dynamically stable radius, can
be used for the parameterization of the spatial distribution of stars
in different brightness ranges. Based on this assumption, we propose
here a method according to which stellar stratification in a star
cluster can be quantified directly by estimating the corresponding
Spitzer radius for every group of magnitudes, using
Eq. (\ref{eq-01}). This simple approach provides information on both
{\sl where} in the cluster every stellar-luminosity group is confined,
and at {\sl which brightness} stellar stratification occurs.

\subsection{Spitzer Radius and Stellar Stratification}\label{s:stratheory}

We test the hypothesis outlined above on artificial populous spherical
clusters that we constructed based on the use of the Monte Carlo
technique. The proposed method is mainly designed for the identification
of stellar stratification in clusters in the Magellanic Clouds (MCs),
and therefore the simulated clusters are taken to have structural
parameters similar to those in the MCs. Results based on observations of
MCs clusters with {\sl HST} are particularly considered due to their
deepness and spatial resolution. Specifically, a sample of four star
clusters (NGC~1818, NGC~2004, NGC~2100, and NGC~330) observed by Keller
et al. (2000) with the {\sl Wide-Field Planetary Camera 2} (WFPC2) is
used as an initial guideline in our simulations, by considering their
structural parameters for the artificial star clusters we construct
here. As a consequence, our simulated clusters have tidal radii of
$r_{\rm t}\simeq$ 2\farcm0, while King models with $C=\log{r_{\rm
t}/r_{\rm c}} \sim 1.0$ are assumed to represent their density profiles.
The corresponding core radii, $r_{\rm c}$, of the artificial clusters
have values within the limits given by Mackey \& Gilmore (2003a,b), who
compiled two-color {\sl HST} observations for a sample of 53 and 10 rich
star clusters in the Large and Small Magellanic Cloud (LMC, SMC),
respectively.

\begin{figure*}[t!]
\epsscale{1.15} 
\plotone{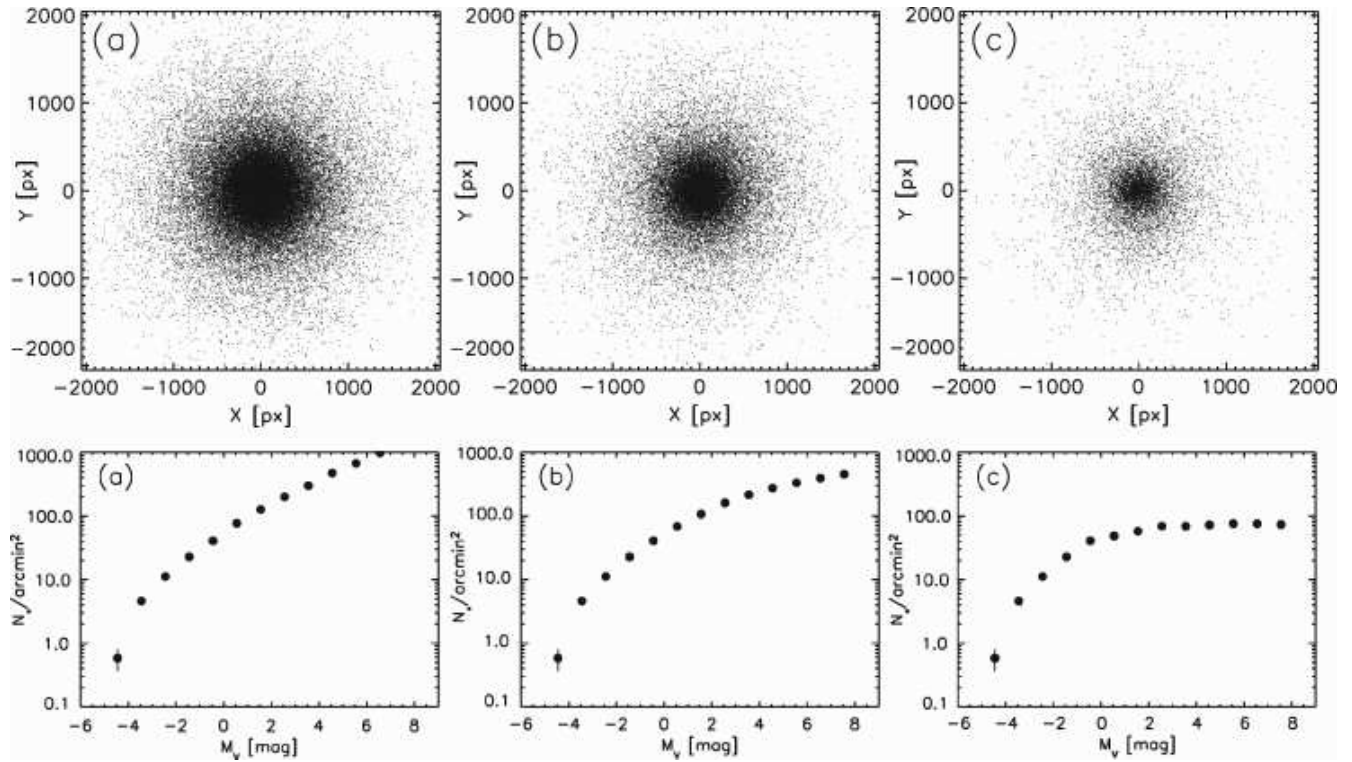}
\caption{Sample of three different simulated star clusters using Monte
Carlo simulations. The maps of the star clusters, which differ in
their luminosity functions, are shown in the top panel. The
corresponding LFs are shown in the bottom panel. All clusters were
assumed to follow a King density profile and to have the same tidal
radius of 0\farcm2. They are also similarly segregated. The LFs shown
are in agreement with the results of various authors on star clusters
in the MCs.\label{f:mcsample}}
\end{figure*}

The LFs of the simulated clusters are selected as variations of a
global LF, which was constructed based on the observed LFs of the
clusters in the sample of Gouliermis et al. (2004). This global LF is
found to be in very good agreement with the LFs of star clusters in
the MCs as determined using {\sl HST} imaging by other authors (e.g., de
Grijs et al. 2002c: NGC~1805, NGC~1818, NGC~1831, NGC~1868, NGC~2209,
and Hodge~14; Sirianni et al. 2002: NGC~330; Testa et al. 1999,
Brocato et al. 2003: NGC~1866; Santiago et al. 2001: NGC~1805,
NGC~1818, NGC~1831, NGC~1868, and NGC~2209). An example of three
artificial clusters with different LFs is given in
Fig.~\ref{f:mcsample}. The LF in each case is also given at the bottom
panel of the figure. All three clusters have the same tidal radius and
the same degree of stratification, but the differences in their LFs
are fed through to their stellar numbers and consequently to their
appearance.

\begin{figure}[t!]
\epsscale{1.15} 
\plotone{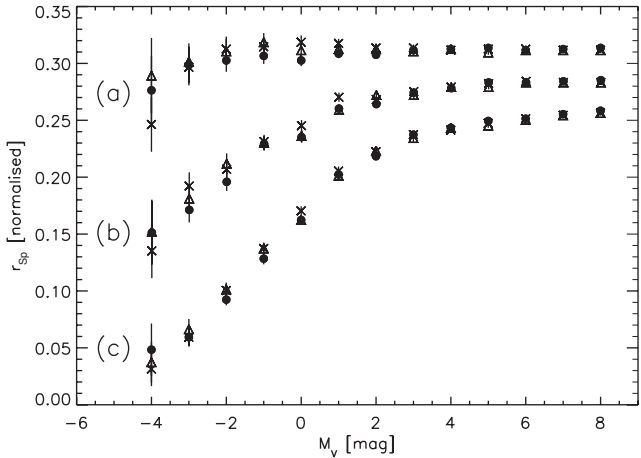}
\caption{Sample of three different degrees of segregation: (a) No
stellar stratification, (b) smooth stratification, and (c) for more
strongly segregated bright stars. The artificial cluster shown in the
middle panel of Fig. \ref{f:mcsample} is used for this demonstration.
The brightest (segregated) stars appear to be confined within smaller
Spitzer radii than the faint (non-segregated) stars.  Different
symbols for each magnitude range represent $r_{\rm Sp}$ as it would be
estimated if the cluster were observed projected onto the $xy$, $xz$,
or $yz$ plane, respectively.\label{f:rspitzsmpl}}
\end{figure}

Stellar stratification was considered for our artificial star
clusters.  Hence, several degrees of segregation (including no
segregation) were adopted. This was done using a projected limiting
radius within which each stellar group should be confined. Based on
our hypothesis, stellar stratification in these clusters should be
detected through the dependence of the estimated Spitzer radius of
stars in specific magnitude ranges versus the corresponding mean
magnitude. Indeed, our tests show that the initially assumed stellar
stratification (of any degree) is reconstructed by the plot of Spitzer
radii per magnitude range versus magnitude for all artificial star
clusters considered. In Fig.~\ref{f:rspitzsmpl} we show a sample of
three degrees of stratification for the same simulated cluster
(similar to the middle panel in Fig.~\ref{f:mcsample}). Spitzer radii
in this figure are normalized for reasons of comparability. Three
different symbols were used for the plot, as if the cluster was
observed projected onto the $xy$, $xz$, or $yz$ plane, respectively.

\subsection{Spitzer Radius of Observed Star Clusters}

Identical to the concept discussed in the previous sections,
observations of the radial distribution of stars in a real star cluster
give us a relation of the kind of Eq. (\ref{eq-01}), \beq r_{\rm
obs}=\sqrt{\frac{\displaystyle\sum^{N_{\rm obs}}_{i=1}
{r_{i}}^{2}}{N_{\rm obs}}} . \label{eq-02} \eeq However, this
distribution is significantly affected by two important observational
constraints: (i) incompleteness of the stellar sample and (ii)
contamination by field stars. To obtain an accurate measurement of the
observed Spitzer radii per magnitude range in a cluster, both these
constraints should be considered.

\subsubsection{Incompleteness of the observations}

The observed stellar samples in star clusters are incomplete due to the
observations. Although this can lead to a parameterization significantly
differing from reality, the bias introduced to the data analysis is, in
general, well understood. Consequently, solving the problem of
incompleteness is a more or less straightforward procedure, which takes
place through extensive artificial-star tests (e.g. de~Grijs et al.
2002a; Gouliermis et al. 2004). A set of artificial stars is generated
within each of the observed frames.  Then, an identical reduction
procedure is performed on the artificially enriched frames, in order to
estimate the number of the `recovered' artificial stars. The
completeness factor $C$ is the ratio between the number of stars
recovered to the number of stars originally simulated.  This factor
depends on the brightness of the stars and their positions in the
cluster. Therefore, an effective completeness estimation for a stellar
cluster requires that incompleteness is calculated for stars in
different magnitude bins and for different distances from the center of
the cluster.

In order to correct the observed Spitzer radius for incompleteness, for
every counted star its completeness factor is calculated according to
its magnitude and radial distance. If the radial distance of the $i^{\rm
th}$ star is $r_{i}$ and its corresponding completeness factor is
$C_{i}$ then we assign $1/C_{i}$ stars to this distance and the total
number of counted stars is affected accordingly. Thus, the
completeness-corrected observed Spitzer radius should be \beq r_{\rm
obs,cc}=\sqrt{\frac{\displaystyle\sum_{i=1}^{N_{\rm
obs}}({r_{i}}^{2}/C_{i})} {\displaystyle\sum_{i=1}^{N_{\rm
obs}}(1/C_{i})}} . \label{eq-03} \eeq

\subsubsection{Contribution from the field populations.}

The contamination of the observations by the field populations is
probably the most difficult problem to be dealt with. Theoretically,
if even one massive field star was counted as a cluster member and if
it is far away from the center of the cluster, then the measured
radius would be overestimated. In the same context, with comprehensive
observations of the field we know the expected number of contaminating
stars per magnitude range, but we still cannot know their position,
which is the most important constraint as far as the radius estimation
is concerned.

One may confront this problem by first defining the largest radial
distance among the stars of the entire sample in a specific magnitude
bin and then normalizing the known surface density of field stars in
this magnitude range to the surface confined by this radius. Hence,
one could obtain a good estimation of the number of field stars
expected to contaminate this area in the specific magnitude range, and
correct for field contribution by subtracting the appropriate fraction
of the total number of field stars. If, for example, there are $N_{\rm
F}$ field stars expected in a total number, $N$, of stars in a
specific magnitude range (within a maximum radial distance, $r_{\rm
max}$), then we can subtract their fraction $F = N_{\rm F}/N < 1$ from
every counted star with distance $r_{\rm i} \leq r_{\rm max}$.  Then,
the field-subtracted observed Spitzer radius, ignoring for the moment
any incompleteness corrections, derived from Eq.  (\ref{eq-02}) for
stars in every magnitude range should be
\beq 
 r_{\rm obs,fs} = \sqrt{\frac{\displaystyle\sum_{i=1}^{N_{\rm obs}} \left[
(1 - {F}){r_{i}}^{2}\right]}{N_{\rm obs} - N_{\rm F}}} . \label{eq-04}
\eeq

However, this simplified approach assumes that the field stars are
more or less homogeneously distributed within the boundaries of the
cluster, which does not correspond to reality. Specifically,
Eq.~(\ref{eq-04}) weighs the contribution of each star to the sum of
${r_{i}}^{2}$ by the probability, $(1-F)$, that it is a cluster
member.  Still, cluster members are by their very nature more
centrally concentrated, and therefore stars at smaller radii are less
likely to be field stars, while those at large radii have a higher
probability that they are, in fact, field stars. As a consequence, in
Eq.~(\ref{eq-04}) true cluster members make a smaller contribution to
the Spitzer radius than they should, while the contribution of the
field stars is overestimated. We verified this problem using
additional Monte Carlo simulations for the construction of
hypothetical background-field populations. For these simulations, the
use of a specific field LF was the only constraint considered.


We constructed homogeneously distributed stellar populations with LFs
typical for the background fields of both Magellanic Clouds, based on
{\sl HST} observations (e.g., Elson et al. 1997; Holtzman et al. 1999;
Dolphin et al. 2001; Smecker-Hane et al. 2002). Subsequently, we
inserted our artificial star clusters (\S~\ref{s:stratheory}) into the
artificial background field. We constructed 32 different cases of four
different clusters, each having four different degrees of stratification
located in two different types of field (with different LFs). The
application of our method to these artificial observations showed that
{\sl it is indeed very difficult to disentangle the contribution of the
field} in the measured Spitzer radii on the basis of Eq.~(\ref{eq-04}).
Consequently, and since it is not possible to quantify the effect of the
field contribution to observed stellar samples in clusters as a function
of distance from the cluster center, a more robust approach for the
direct field decontamination of the observed stellar samples {\sl before
measuring the Spitzer radii} should be considered. Such a
well-established method for the removal of any field contamination from
the stellar samples of star clusters is based on the application of a
Monte Carlo technique, which makes use of comprehensive observations of
the local background field of the galaxy (e.g., Bonatto \& Bica 2007).
We apply such a sophisticated technique to actual {\sl HST}/ACS
observations of two SMC clusters, for the estimation of their Spitzer
radii as a function of stellar magnitude, in \S~3.


\subsection{The Effective Radius}

The method we propose for the diagnosis of stellar stratification in
star clusters makes use of the observable counterpart of Spitzer
radius, applied to the observed completeness-corrected stellar
samples, and after the contamination of these samples by the field
populations has been removed. This radius, which we will refer to as
the {\sl effective radius}, $r_{\rm eff}$, is estimated for stars in
specific narrow magnitude ranges; its dependence on magnitude is shown
in the following sections. Eq. (\ref{eq-03}) can be reduced to the
following expression and should, with all corrections properly
applied, be equal to $r_{\rm Sp}$. Thus, \beq r_{\rm
eff}=\sqrt{\frac{\displaystyle\sum_{i=1}^{N_{\rm
obs,cc,ff}}{\frac{{r_{i}}^{2}}{C_{i}}} }{N_{\rm obs,cc,ff}}} \equiv
r_{\rm Sp} \label{eq-07} , \eeq where $r_{i}$ is the radial distance
of the $i^{\rm th}$ stellar member of the cluster (after field
subtraction), in a specific magnitude range. $C_{\rm i}$ is the
corresponding completeness factor at its distance for the same
magnitude range and $N_{\rm obs,cc,ff}$ is the total number of
observed stars corrected for incompleteness and reduced by the total
number of field stars.

This radius is a statistical quantity and its estimation is rather
sensitive to the number of observed stars. Therefore, the effective
radius can be easily over- or underestimated if this estimation is
based on few stars only, which might be the case for the brightest
magnitude ranges. In order to deal with this issue, one may apply the
binning in variable magnitude ranges and use wider magnitude bins
toward the brighter limit, thus increasing the number of the sample
stars. However, with this solution one may not be able to observe the
luminosity segregation of the stars in detail. It would be more
appropriate to weigh each bin by the number of sample stars as an
indicator of the statistical significance of the estimation of the
corresponding effective radius.

The uncertainty in the estimation of $r_{\rm eff}$, as derived from
Eq. (\ref{eq-07}), is mostly dependent on errors in the counting
process, thus reflecting Poisson statistics. Another concern to be
taken into account is the effect of projection on the estimated values
for the effective radius. For non-symmetrical loose stellar systems,
it is found that projection does not significantly affect the true
value of the Spitzer radius (e.g., Gouliermis et al. 2000). Here,
since we deal with spherically symmetric clusters, we consider the
uncertainties introduced by projection for the simple case of a
homogeneous density distribution.

\begin{figure}[t!]
\epsscale{1.15} 
\plotone{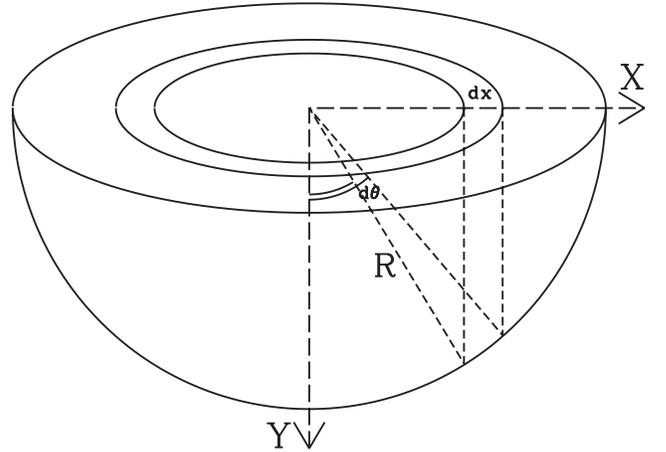}
\caption{Projection of a hemispherical distribution. The $Y$ axis coincides 
with the line of sight.\label{f:projfig}}
\end{figure}

We assume a spherical distribution of $N$ particles ($N \gg 1$),
uniformally distributed (with constant density, $\varrho$) within a
radius, $R$. If one observes this sphere, the distribution is not
uniform anymore because of projection effects.  Specifically, we
investigate the size of the radius which includes half of the
particles, projected along the line of sight, in relation to the
radius of a sphere which -- in reality -- includes half of them (in 3D
space). Because of symmetry we concentrate on one hemisphere, defined
by the line of sight (direction of the Y axis in
Fig.~\ref{f:projfig}).

We define an elementary cylinder of thickness $dx =
R~\cos{\theta}~d\theta$, of base radius $x = R~\sin{\theta}$, and
of height $y = R~\cos{\theta}$ (Fig.~\ref{f:projfig}). The angle
$\theta$ is measured in relation to the radius vertical to the surface
of projection. The cylinder contains $dN = \varrho~dV$
particles, where $V$ is the volume. Consequently,
\beq
dN\left ( \theta \right ) = 2\pi~R^{3}~\varrho~\sin{\theta}~ 
\cos^{2}\mkern-4mu{\theta}~d\theta .
\eeq

Integrating over $0{\deg} < \theta < 90{\deg}$ we expect that the
number of particles included is $N = 2\pi~R^{3}~\varrho/3$. Half
($N/2$) are projected up to a limit cylinder base radius, $x_{\rm h}$,
corresponding to $\theta_{\rm h}$. This angle can be deduced from the
relation \begin{displaymath} N/2 = \int_{0}^{\theta_{\rm h}} {\rm
d}N\left ( \theta \right) = -2 \pi R^{3} \varrho \int_{0}^{\theta_{\rm
h}} \cos^{2}\mkern-4mu{\theta}~d \left ( \cos{\theta} \right )
~\Rightarrow \end{displaymath} \beq \Rightarrow
~\cos^{3}\mkern-4mu{\theta_{\rm h}} = \frac{1}{2} ~\Rightarrow~
\theta_{\rm h} \approx 37\farcd467 . \eeq This angle gives $x_{\rm h}
= R~\sin{\theta_{\rm h}} \approx 0.608R$. In reality, half of the
particles are included in a volume contained within a sphere of radius
$R_{\rm h} = 2^{-1/3}R \simeq 0.794R$ (because of constant
density). Thus, we have $R_{\rm h} = 1.3x_{\rm h}$.  As a consequence,
the proportion of the half-number radius (and consequently of the
Spitzer radius, which is almost equal to the half-number radius for a
relaxed stellar system) to its projected value for this distribution
of particles is well-defined.

\begin{figure}[t!]
\epsscale{1.15} 
\plotone{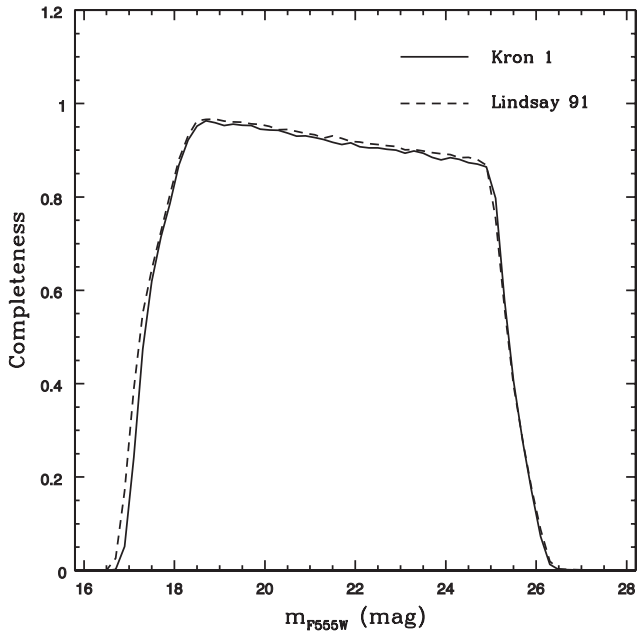}
\caption{Completeness functions for Kron~1 (solid line) and Lindsay~91
(dashed line), respectively. The function has been integrated over
position and color as a function of m$_{\rm F555W}$.\label{f:compl}}
\end{figure}

In any case, in the present study this factor is irrelevant for
further analysis, since the application of a constant conversion
factor to the estimated effective radii is not necessary. Moreover,
this application assumes that any cluster follows a homogeneous
density distribution, which is of course not always the case. Under
these circumstances, we can treat the estimated effective radii used
in the application of our method as the {\sl projected} effective
radii.

\begin{figure*}[t!]
\epsscale{1.15} 
\plotone{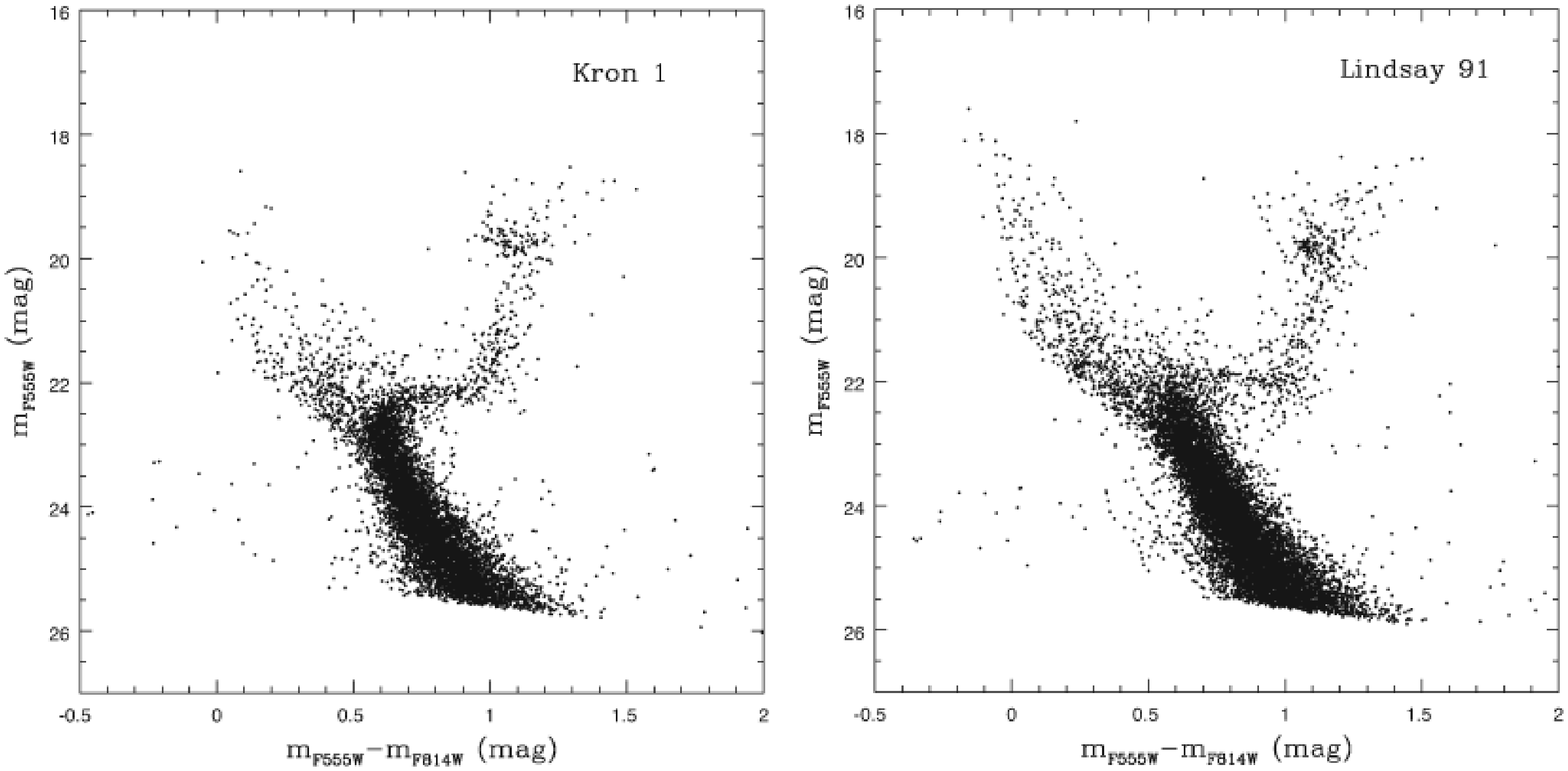}
\caption{m$_{\rm F555W}$$-$m$_{\rm F814W}$ vs. m$_{\rm F555W}$ CMDs of 
the fields of the SMC clusters Kron~1 and Lindsay~91, observed with
{\sl HST} ACS/WFC.\label{f:acscmds}}
\end{figure*}

\section{The Star Cluster Sample}\label{s:data}

The proposed method for the detection of stellar stratification is
designed primarily for the study of MC clusters, and therefore in the
following sections we apply this method to such clusters to test its
performance and to establish a consistent methodology. Moreover, in
order to use the most complete available data we select clusters
observed with the {\sl HST}. The advantages {\sl HST} introduced to
crowded-field photometry in the MCs have been documented extensively
by several authors since its early observations and the quality of the
data obtained with both the Wide-Field Planetary Camera 2 (WFPC2) and
the Advanced Camera for Surveys (ACS) has proven to be more than
adequate for detailed studies of young and old MC star clusters (e.g.,
Fischer et al. 1998; Johnson et al. 1999; Keller et al. 2000; Brocato
et al. 2001; de~Grijs et al. 2002a,b,c; Stanghellini et al. 2003;
Gouliermis et al. 2004; Mackey et al. 2006; Rochau et al. 2007; Xin et
al. 2008).

For the application of our new method for diagnosis of stellar
stratification to real clusters we selected a set of data obtained with
the ACS of the intermediate-age clusters Kron~1 (Kron 1956) and
Lindsay~91 (Lindsay 1958) in the SMC. Below, we discuss the reduction of
the observations of these clusters (see also Mackey \& Gilmore 2004),
their photometry, and the decontamination of the cluster populations
from the contribution of the general background field of the galaxy
using an advanced Monte Carlo technique.

\subsection{Data Reduction}

The SMC clusters Kron~1 and Lindsay~91 were observed with the Wide
Field Channel (WFC) of ACS on-board the {\sl HST} (program GO-9891, PI
G. Gilmore). The ACS/WFC consists of two 2048$\times$4096-pixel CCDs,
separated by a gap of $\sim$50 pixels.  It covers a field of view of
202$\times$202 arcsec$^2$, with a scale of $\sim$0.05 arcsec
pixel$^{-1}$.  The frames were taken in each of the F555W and F814W
filters.  Exposure times were 300 and 200 seconds, respectively. More
details on the observations and instruments can be found in Mackey \&
Gilmore (2004).

The {\sc fits} files of the two clusters were retrieved from the Space
Telescope Science Institute (STScI) Data Archive. The original data
were reduced with the STScI pipeline, i.e., they had been bias and
dark-current subtracted and divided by flat-field images.  Photometry
was performed with the ACS module of the package {\sc
dolphot}\footnote{The ACS module of {\sc dolphot} is an adaptation of
the photometry package {\sc HSTphot} (Dolphin 2000). It can be
downloaded from {\tt http://purcell.as.arizona.edu/dolphot/}.}
(Version 1.0). Photometry processes and corresponding parameters fully
follow the procedures and recommendation in the {\sc dolphot} manual.
Photometric calibrations and transformations were done using the
relations in Sirianni et al. (2005).

We adopted three parameters in {\sc dolphot} to filter the photometric
results, i.e., we selected only objects with
$-$0.3~$\leq$~sharpness~$\leq$~0.3, crowding~$\leq$~0.5~mag, and
$\chi^2$~$\leq$~0.25 in both frames. Meanwhile, we kept objects
classified as good star (type 1) and star errors of types 1 to 7 by {\sc
dolphot}, which are referred to as `usable' in the {\sc dolphot} manual.
To calculate the completeness of the photometry, {\sc dolphot} was run
again in the artificial-star mode. For each cluster, we generated
$\sim10^6$ fake stars with limits of 16.0~-~28.0~mag in brightness and
$-$0.50~-~2.00~mag in color. The effect of unrealistically large crowding
that would lead to over-estimation of the incompleteness was taken into
account by the {\sc dolphot} utility {\sl acsfakelist}, which creates
artificial star lists based on the original photometric catalog. The
fake stars were binned in four dimensions, i.e, $x$ and $y$ positions,
magnitude, and color. Fig.~\ref{f:compl} presents the completeness
function of Kron~1 (solid line) and Lindsay~91 (dashed line),
respectively. The function is integrated over position and color and is
a function of m$_{\rm F555W}$.  More details on the data reduction are
given in Xin et al. (2008).

\subsection{Field Decontamination}\label{ss:fd}

\begin{figure*}[t!]
\epsscale{1.15} 
\plotone{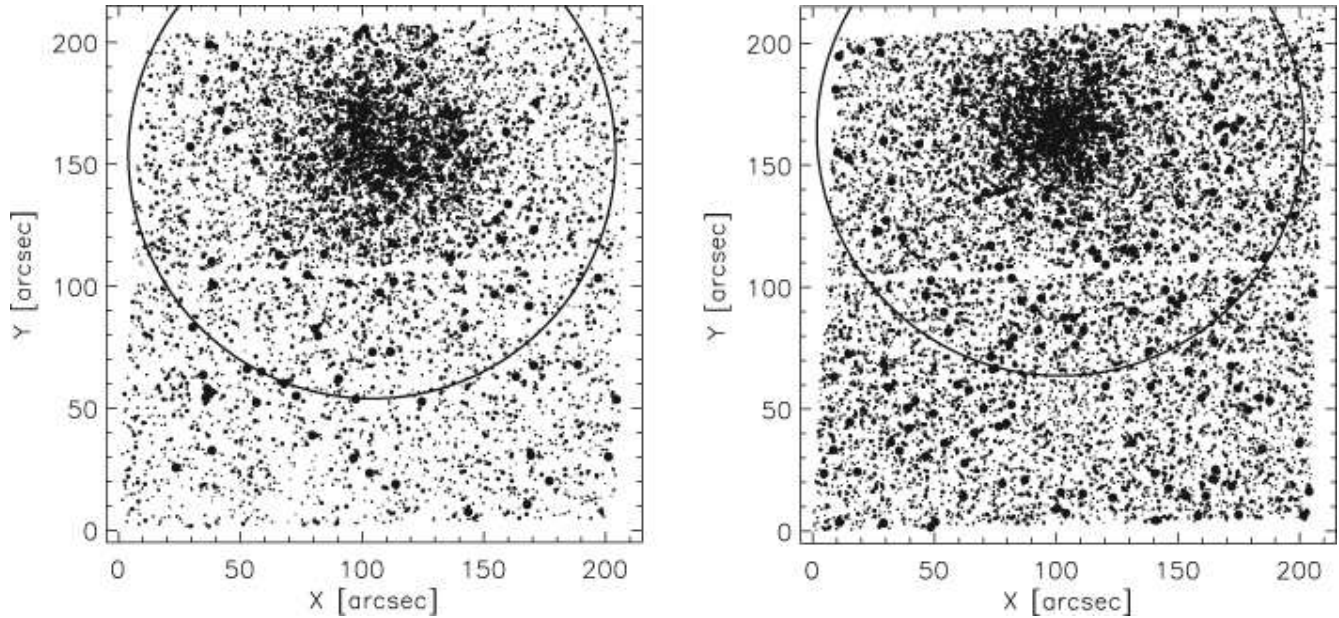}
\caption{The observed ACS/WFC fields of view of Kron~1 (left) and
Lindsay~91 (right).  The overlayed circles have radii, $R_{\rm lim}$,
corresponding to the boundary between the clusters and the general
field. They were defined based on the clusters' surface density
profiles, as the distance from the cluster center where the density
profiles become flat (see \S~\ref{ss:fd}).\label{f:acsmaps}}
\end{figure*}

The observed color-magnitude diagrams (CMDs) of the ACS fields centered
on Kron~1 and Lindsay~91 are shown in Fig.~\ref{f:acscmds}. Both fields
suffer from significant contamination by field stars, as can also be
seen in the corresponding maps shown in Fig.~\ref{f:acsmaps}. Without
any information about cluster-membership probabilities for the observed
stars, obtained, e.g., from radial velocities and/or proper motions, the
quantitative decontamination of the clusters from the field stars on a
statistical basis becomes a fundamental method to obtain the CMD of the
true stellar members of each cluster (see e.g. the pioneering work by
Flower et al. 1980 on the LMC cluster NGC~1868).

In this work, the algorithm described in Bonatto \& Bica (2007) is
used for field-star decontamination of the observed CMDs of both
clusters.  As a first step, we use the stellar number-density profiles
of the clusters to identify the radial distance from the center of
each cluster, R$_{\rm lim}$, where the stellar density becomes flat,
indicating that the background-field density has been reached. We call
this region the `offset region' of the cluster. All stars in the
offset regions (with $r>R_{\rm lim}$) are treated as field stars,
while the rest (with $r \leq R_{\rm lim}$) are considered as the most
probable cluster-member stars.  Assuming a homogeneous field-star
distribution, the number density of field stars is applied to the
whole cluster region to remove the field contamination. We perform
this calculation in two dimensions in the CMD, i.e., in m$_{\rm
F555W}$ and in (m$_{\rm F555W}$$-$m$_{\rm F814W}$), considering also
the observational uncertainties in the photometry, $\delta_{\rm
F555W}$ and $\delta_{\rm F814W}$.

In short, the decontamination process is done as follows: (i) we
divide the CMDs of the cluster and offset-region stars, respectively,
in 2D cells with the same axes along the m$_{\rm F555W}$ and (m$_{\rm
F555W}$$-$m$_{\rm F814W}$) directions, (ii) we then calculate the
expected number density of field stars in each cell in the CMD of the
offset region, and (iii) we randomly subtract the expected number of
field stars from each cell from the CMD of the cluster region. In the
following description we use the symbols $\chi$=m$_{\rm F555W}$ and
$\xi$= (m$_{\rm F555W}$-m$_{\rm F814W}$) to simplify the
notation. Consider a CMD cell with sides and coordinates
$\chi_c\pm\triangle\chi/2$ and $\xi_c\pm\triangle\xi/2$, respectively,
where ($\chi_c$, $\xi_c$) are the cell's central coordinates. We
assume a Gaussian probability distribution to calculate the
probability of a star with CMD coordinates
($\bar{\chi}\pm\delta_\chi$, $\bar{\xi}\pm\delta_\xi$), with
$\delta_\xi$=($\delta^2_{\rm F555W}$+$\delta^2_{\rm F814W}$)$^{1/2}$,
to be found in the cell, and therefore the computations take into
account both magnitude and color uncertainties, for example,

\beq
P(\chi,\bar{\chi})=\frac{1}{\sqrt{2\pi}\delta_\chi}e^{(-1/2)\left[(\chi-\bar{\chi})/\delta_\chi\right]^2} .
\eeq

The expected field-star number density ($\rho^{\rm cell}_{\rm fs}$) in
a cell is given by summing up the individual probabilities ($P^{\rm
cell}_{\rm fs}$) of all offset-field stars (N$_{\rm fs}$) in the cell,
divided by the offset area (A$_{\rm fs}$), i.e., $\rho^{\rm cell}_{\rm
fs}$=$P^{\rm cell}_{\rm fs}$/A$_{\rm fs}$, where

\beq
P_{\rm fs}^{\rm cell}=\sum_{i=1}^{N_{\rm fs}}\int\int P_i(\chi, \chi_i; \xi, \xi_i) d\chi d\xi .
\eeq

$P_i(\chi, \chi_i; \xi, \xi_i)$ is the probability of the
\textit{i}$^{\rm th}$ field star, with CMD coordinates ($\bar{\chi}$,
$\bar{\xi}$) and uncertainties ($\delta_{\chi i}$, $\delta_{\xi i}$), to
have the magnitude and color ($\chi$, $\xi$).  The integral is carried
out in the two dimensions, $\chi_c$: $\triangle \chi/2 \leq \chi \leq
\chi_c + \triangle \chi/2$, and $\xi_c$: $\triangle \xi/2 \leq \xi \leq
\xi_c + \triangle \xi/2$, respectively.  In the same manner we calculate
the number density of the observed stars in the cell in the
cluster-region CMD, $\rho^{\rm cell}_{\rm obs}=P^{\rm cell}_{\rm
obs}$/A$_{\rm cl}$, where A$_{\rm cl}$ is the projected area of the
cluster region ($r \leq R_{\rm lim}$). Therefore, the expected number of
field stars in the cell in the cluster-region CMD is $n^{\rm cell}_{\rm
fs} = (\rho^{\rm cell}_{\rm fs}/\rho^{\rm cell}_{\rm obs}) \times n^{\rm
cell}_{\rm obs}$, where $n^{\rm cell}_{\rm obs}$ is the number of
observed stars (at $r\leq R_{\rm lim}$) located in the cell. The number
of probable cluster member stars in the cell will be $n^{\rm cell}_{\rm
cl} = n^{\rm cell}_{\rm obs}-n^{\rm cell}_{\rm fs}$, and the total
number of cluster member stars is the sum of all $n^{\rm cell}_{\rm
cl}$, i.e., $N_{\rm cl}=\sum_{\rm cell}n^{\rm cell}_{\rm cl}$. 

To minimize any artificial effect intrinsic to the method, we apply the
algorithm many times using different cell sizes ($\triangle\chi$,
$\triangle\xi$) and derive different decontamination results, from which
we calculate the probability of a star to be identified as a `true'
cluster member. Finally the field-star-decontaminated CMD of the
cluster contains the $N_{\rm cl}$ stars with the highest probability of
being cluster members.  Fig.~\ref{f:fsdcmds} shows the corresponding
CMDs of the cluster region (left panel), the offset region (central
panel), and the final field-star-decontaminated CMD (right panel) for
both clusters.

\section{Application of the diagnostic method}\label{s:application}

The proposed method for the diagnosis of stellar stratification is
based on the assumption that the {\sl effective radius} of stars in a
specific magnitude (or mass) range will be a unique function of this
range if the system is segregated. In this section we apply our new
method for the detection and quantification of stellar stratification
to the intermediate-age clusters Kron~1 and Lindsay~91. Before the
application of our diagnostic method and to check if the clusters are
segregated, we investigate the stellar stratification using one of the
most accurate `classical' methods, the radial dependence of the
cluster LFs and MFs.

\begin{figure*}[t!]
\epsscale{1.15} 
\plotone{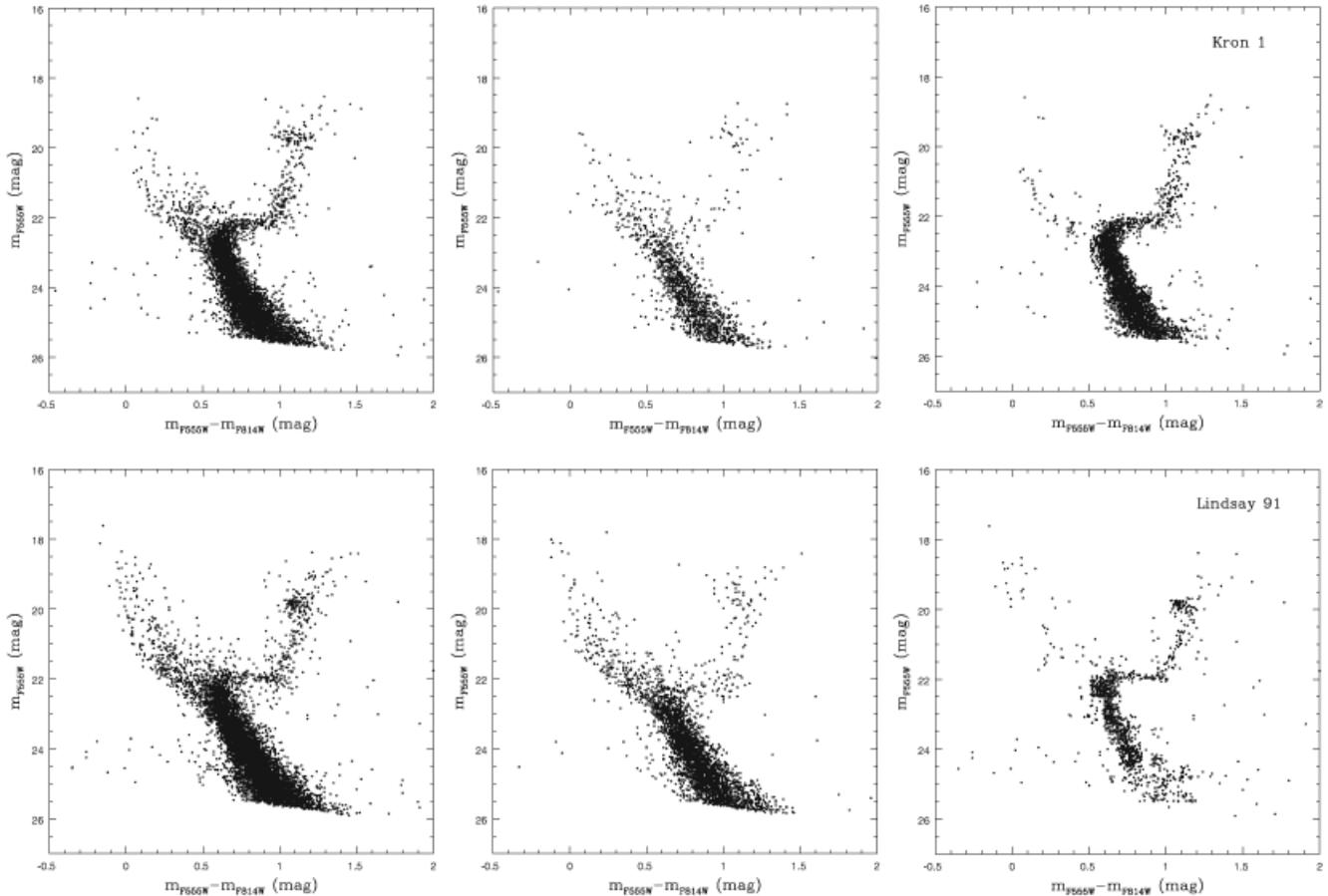}
\caption{CMDs of the cluster regions ($r \leq R_{\rm lim}$; left panel),
the offset regions ($r > R_{\rm lim}$; central panel), and the final CMDs
of the clusters after field-star decontamination (right panel) for both
Kron~1 (top) and Lindsay~91 (bottom).\label{f:fsdcmds}}
\end{figure*}

In the previous section we showed that both clusters are evolved, by
virtue of the presence of clear red-giant branches (RGBs) and red
clumps in their CMDs. However, in the corresponding `clean' CMDs,
obtained after decontamination of the cluster stellar samples by the
field populations, shown in Fig.\ref{f:fsdcmds} (right panels), it can
be seen that there are stars at the upper main sequence (MS) and just
above the MS turn-off, which are not sufficiently removed from the
observed CMDs. These stars are most probably members of the younger
SMC field, to which both clusters belong, and they will not be
considered in the following analysis. As a consequence, for the
investigation presented here, we use only the lower-MS population
below the turn-off of each cluster and that of the RGB.

\subsection{Stellar Stratification in the selected clusters}

In this section we apply the most effective {\sl classical} diagnostic
method developed in the past for the investigation of stellar
stratification to our ACS photometry of Kron~1 and Lindsay~91, to
check if the clusters are indeed segregated. This well-applied method
requires the construction of the cluster LF and/or the MF and the
investigation of its radial dependence. If the cluster is indeed
segregated then the appearance of more massive stars toward its center
will result in a shallower MF at smaller radii. De Grijs et al.
(2002b) present the difficulties in the construction of an accurate MF
based on the use of different ML relations and stress the differences
in the MF slopes obtained. Moreover, significant numerical biases in
the determination of the slope of the MF using linear regression are
found if the construction of the MF was made from uniformly binned
data (Ma{\'{\i}}z Apell{\'a}niz \& {\'U}beda 2005), implying the
presence of systematic errors in the slopes of MFs calculated in this
way.

Gouliermis et al. (2004) argue that the investigation of any radial
dependence of the MF slope is not as straightforward as it seems,
since any such dependence is rather sensitive to the selection of the
radial distances being considered. Although stellar stratification can
be identified from the cluster LF, its quantification is quite
difficult because the approximation of the LF with a single power law,
which is the usual approach, can only give very rough results. In any
case, we apply this diagnostic to our data, in order to identify
stellar stratification in our clusters and to be able to compare the
results of a well-established classical method with those of our
method.

\subsubsection{Radial Dependence of LFs and MFs of the Clusters}

\begin{figure*}[t!]
\epsscale{1.15} 
\plotone{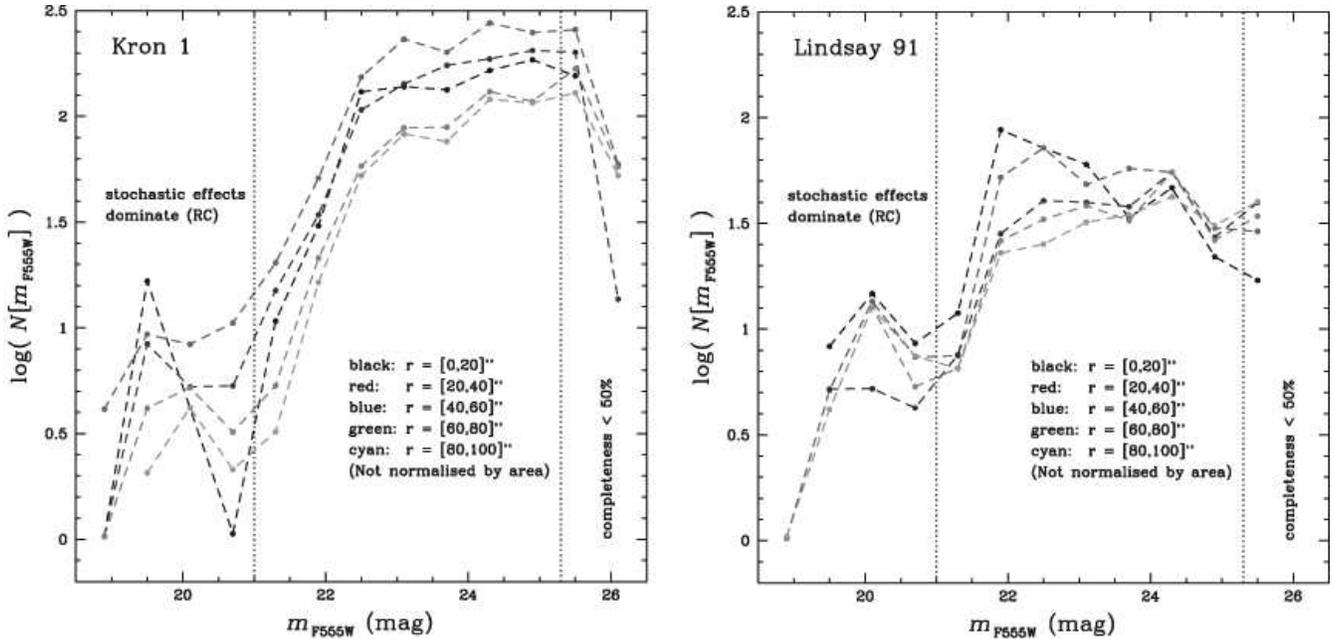}
\caption{Completeness-corrected LFs of Kron~1 (left) and Lindsay~91
(right) within different annuli around the center of the
clusters. These LFs clearly show how difficult it is to derive a
single slope for the entire LF to compare the results for different
clusters. Specific parts of the LF can, however, be approximated by a
single power law. The LF of Kron~1 does not show any prominent
dependence on the radial distance. Lindsay~91, however,
shows a clear radial dependence of its LF in the magnitude range
22.0~\lsim~$m_{\rm F555W}$/mag~\lsim~25.0. One should only consider
the LFs in the magnitude range in between the vertical dotted lines,
as the brightest stars are affected by small-number statistics and the
presence of a red clump (RC), while the lowest-luminosity stars suffer
from significant statistical incompleteness. \label{f:lfs}}
\end{figure*}

In this section we examine the dependence of the shape and slope of
the stellar LF on position within the clusters. We basically use the
cluster LFs as smoothing functions of the full 2D ($m_{\rm F555W},
m_{\rm F555W} - m_{\rm F814W}$) CMDs shown in the right-hand panels of
Fig. \ref{f:fsdcmds}. In Fig. \ref{f:lfs} we show all annular cluster
LFs out to $r = 100.0''$, corrected for the effects of incompleteness
and background contamination at each respective annular radius. We
carefully considered the optimal radial ranges to be used for our LF
analysis, which was in essence driven by the need to have
statistically significant {\it and} similar (cf. Ma{\'{\i}}z
Apell{\'a}niz \& {\'U}beda 2005) numbers of stars in each of the
subsamples used for our comparison of the different annular LFs. For
reasons of clarity, we did not normalize the LFs to the same sampling
area since this would result in smaller separations among the various
curves, hence hampering our assessment of any radial dependence of the
LF {\it shape} (at this point we are not interested in the absolute
numbers of stars per unit area).

We note that for a proper comparison of the annular LFs in Fig.
\ref{f:lfs} one should only consider the magnitude range in between the
vertical dotted lines. For brighter stars, stochasticity in the stellar
LFs starts to dominate, while the LFs are also significantly affected by
the presence of red-clump stars, which are clearly visible in both CMDs.
For fainter stars our subsamples are significantly statistically
incomplete, at $<$50\%. On the basis of a casual comparison of the
annular LFs in Fig. \ref{f:lfs}, we conclude that there is little
evidence for luminosity segregation in Kron 1, although there is a hint
of a flattening of the LFs towards smaller radii, particularly for
$m_{\rm F555W}~\gsim~22.5$ mag. Lindsay 91, on the other hand,
exhibits clear signs of luminosity segregation for most of the magnitude
range of interest.

The conversion of an observational LF in a given passband {\it i},
$\phi (M_i)$, to its associated MF, $\xi (m)$, is not as
straightforward as often assumed (see, e.g., de Grijs et al. 2002b).
The differential present-day stellar LF, d{\it N}/d$\phi (M_i)$, i.e.
the number of stars in the absolute-magnitude interval $[M_i,M_i+{\rm
d}M_i]$, and the differential present-day MF, d{\it N}/d$\xi(m)$, i.e.
the mass in the corresponding mass interval $[m,m+dm]$, are
related through d$N = -\phi(M_i) dM_i = \xi(m) dm$, and
therefore
\begin{equation}
\label{MLrelation.eq}
\phi(M_i) = -\xi(m) {dm \over dM_i} .
\end {equation}

Thus, to convert an observational LF into a reliable MF, one needs to
have an accurate knowledge of the appropriate ML -- or
mass--absolute-magnitude -- relation, d{\it m}/d$M_i$ (for a detailed
discussion, see de Grijs et al. 2002b). In fact, it is the {\it slope}
of the ML relation at a given absolute magnitude that determines the
corresponding mass, which is therefore quite model dependent. This has
been addressed in detail by, e.g., D'Antona \& Mazzitelli (1983), Kroupa
et al. (1990, 1993), Elson et al. (1995) and Kroupa \& Tout (1997).
Given the non-linear shape of the ML relation (de Grijs et al. 2002b)
and the small slope at the low-mass end, any attempt to model the ML
relation by either a polynomial fit or a power-law dependence will yield
intrinsically unreliable MFs (cf. Elson et al. 1995, Chabrier \& M\'era
1997), in particular in the low-mass regime. This model dependence is
clearly illustrated by, e.g., Ferraro et al. (1997) and de Grijs et al.
(2002b), who compared the MFs for their sample clusters derived from a
variety of different ML relations at that time available in the
literature. Nevertheless, for the sake of our discussion on mass
segregation in our two sample clusters, we only need to consistently
apply {\it the same} ML relation to the clusters' annular LFs; any
differences in the resulting MF shapes will then be due to intrinsic
differences in the stellar mass distributions as a function of radius in
the clusters.

In addition, the true structure of the CMD above the MS turn-off is very
complex. Stellar populations of different masses overlap in
color-magnitude space, so that unambiguous mass determination from
isochrone fits, for the few dozen stars populating these areas in each
cluster, is highly model-dependent. On the other hand, while different
evolutionary models differ in terms of absolute calibration of the
mass-luminosity relation, they show a quite good agreement in terms of
relative mass distribution in well established evolutionary phases. In
any case, in a differential analysis such as presented here, the
uncertainties involved in the mass determinations of the evolved stars
are too large and systematic, so that we cannot include these stars in
our MF analysis. Therefore, we used all stars below the MS down to the
50\% completeness limit of the data. After applying the appropriate
distance modulus to obtain absolute magnitudes, for our
luminosity-to-mass conversions we used the updated Padova isochrones
(Marigo et al. 2008; available from {\tt\small
http://stev.oapd.inaf.it/cmd}) for the appropriate {\sl HST} photometric
system, the metallicity and age of the cluster, and a Kroupa (1998) IMF
corrected for the presence of a binary population, as suggested for use
with the web interface. For the metallicity and age of both Kron 1 and
Lindsay 91 we used $Z = 0.001$ (where Z$_\odot = 0.020$) and an age of
6.3 Gyr; their distance moduli are $m-M = 19.34$ mag and $m-M = 19.10$
mag, respectively. These physical parameters are derived from fitting
the isochrone model that best matches the observed features of the CMD
of each cluster.

\begin{figure}[t!]
\epsscale{1.} 
\plotone{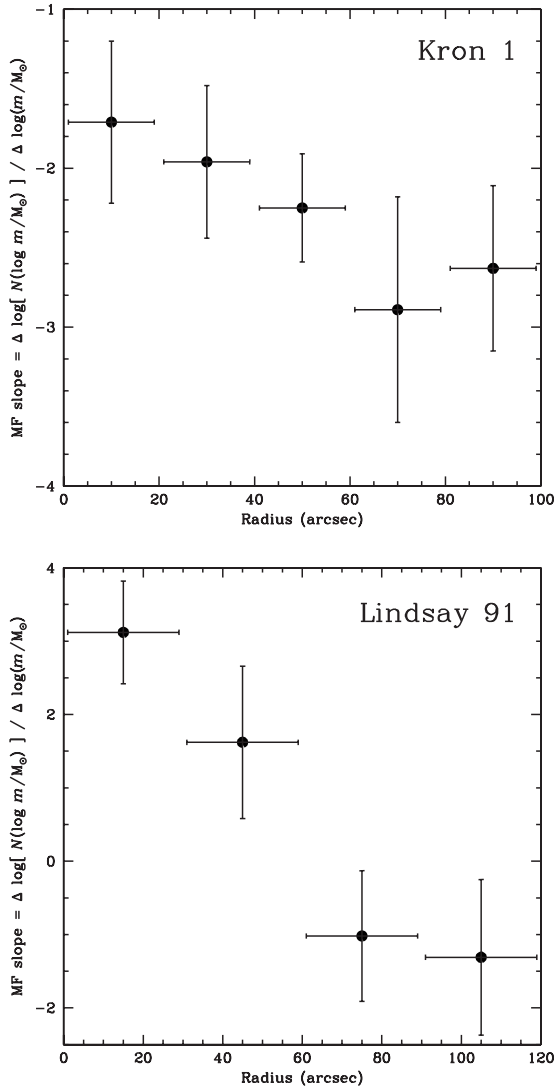}
\caption{Radial dependence of the slopes of the differential MFs of
Kron~1 (top) and Lindsay~91 (bottom). This dependence provides a clear
indication of mass segregation of the stars on the lower MS (below the
turn-off) for both clusters, and particularly for Lindsay 91. However,
the uncertainties in the counting process, the model dependence for
construction of the MF, and the selection of specific radial distances
whithin which the MF is constructed for the analysis of stellar
stratification introduce additional uncertainties, which work their
way through to the results of this analysis.\label{f:mfs}}
\end{figure}

The MS turn-off magnitude in both Kron 1 ($m_{\rm F555W} \simeq 22.3$
mag) and Lindsay 91 ($m_{\rm F555W} \simeq 22.1$ mag) corresponds to a
mass of $\log(m_\ast/{\rm M}_\odot) \simeq -0.009$, while the 50\%
completeness limits correspond to $\log(m_\ast/{\rm M}_\odot) = -0.2$
and $-0.17$ for these clusters, respectively. In Fig. \ref{f:mfs} we
show the derived MF slopes as a function of cluster radius for our
adopted ML conversion in the full mass range; the canonical Salpeter
slope would be $\alpha = -2.35$ in this representation. The adopted
radial ranges for our annular MFs are indicated by the horizontal
`error' bars. The vertical error bars represent the formal uncertainty
in the fits. Although the uncertainties are large, both clusters seem to
be affected by mass segregation, with Lindsay 91 being most obviously
mass segregated, out to at least $r \sim 80''$.
 
\subsection{Effective Radii of the Clusters}

\begin{figure*}[t!]
\epsscale{1.15} 
\plotone{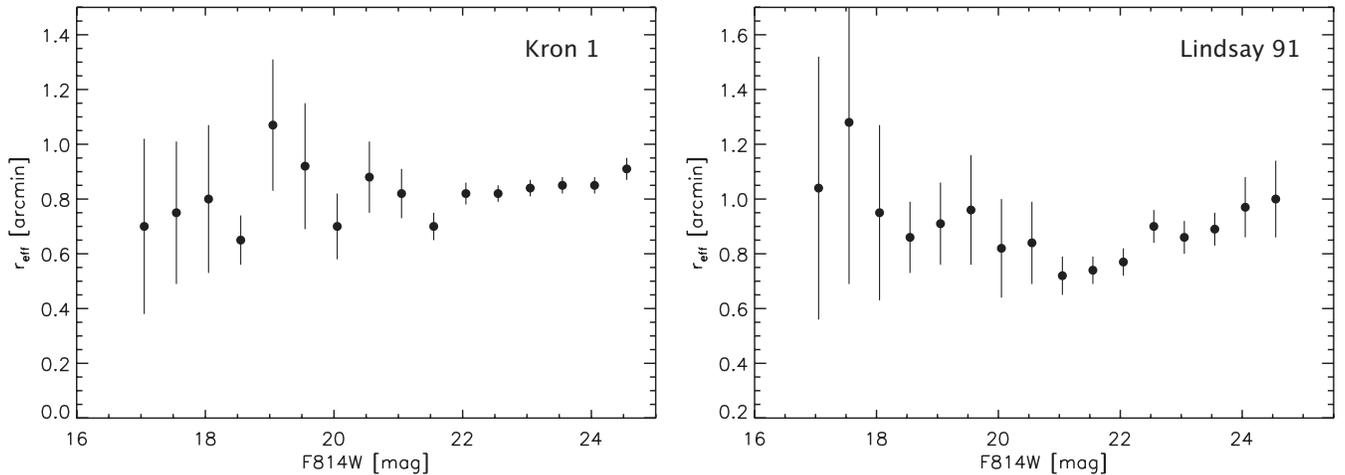}
\caption{Estimated effective radii of stars within different magnitude
ranges versus the corresponding mean magnitude for both
clusters.\label{reffvsmag2}}
\end{figure*}

We calculated the effective radii of the stars in different magnitude
ranges for both clusters. Magnitude bins of various sizes were tested
and we found that a reasonable bin size, which provides good
statistics and allows for the detailed observation of the dependence
of $r_{\rm eff}$ on magnitude, is 0.5~mag. We calculated the effective
radii of the stars as a function of magnitude in the F814W filter. The
relations derived between the computed effective radius and the
corresponding magnitude bin for both Kron~1 and Lindsay~91 are shown
in Fig. \ref{reffvsmag2}. It is shown that, in general, the effective
radius behaves as a function of magnitude for both clusters, thus
providing clear indications of stellar stratification. While this
behavior is more definite for the fainter stars, where it is seen that
for both clusters the effective radii of these stars are larger, the
uncertainties in the $r_{\rm eff}$ calculation for the brightest stars
are rather large due to small-number statistics, and therefore no
definite trend for the bright stars can be derived.

From the $r_{\rm eff}$ vs. magnitude plots in Fig.~\ref{reffvsmag2}
one may conclude that the clusters exhibit stellar stratification but
to a different {\sl degree}. Moreover, from the plots in
Fig.~\ref{reffvsmag2} one can derive the degree of stratification of
the clusters in terms of the brightness range of the segregated stars,
as well as the effective radius within which they are
confined. Considering that a steeper slope of the relation $r_{\rm
eff}(m_{\rm F814W})$ represents a higher degree of stratification, we
conclude that Lindsay~91 is more strongly segregated than Kron~1.

If we want to parameterize the differences between the clusters and
obtain more quantitative results, we should use the two primary output
parameters of our method, i.e., the {\sl magnitude} and the {\sl
effective radius} of segregation.  Both represent the limit (in
magnitude range and radius, respectively) beyond which the slope of
$r_{\rm eff}(m_{\rm F814W})$ changes significantly.  However, as can be
seen in Fig.~\ref{reffvsmag2}, $r_{\rm eff}$ is not a monotonic function
of brightness. Specifically, for Kron~1, while for stars with $m_{\rm
F814W}$~\lsim~19~mag there is a trend of $r_{\rm eff}$ to become larger
for fainter magnitudes, this trend cannot be confirmed statistically due
to large uncertainties. For fainter magnitudes, down to $\sim$~20.5~mag,
the relation $r_{\rm eff}$ vs. $m_{\rm F814W}$ shows fluctuations and no
specific trend. For even fainter stars, with $m_{\rm
F814W}$~\gsim~21~mag, the slope of this relation shows a definite
steepening as a clear indication that these stars are indeed segregated.
However, this slope is still quite shallow, providing evidence of a {\sl
low degree of segregation}. A comparison between the $r_{\rm eff}$ vs.
$m_{\rm F814W}$ relation of Fig.~\ref{reffvsmag2} for Kron~1 and the LFs
of the cluster constructed for different distances from its center
(Fig.~\ref{f:lfs}), shows that $r_{\rm eff}$ seems to be a function of
brightness for the entire observed magnitude range, but statistics do
not allow the verification of a definite relation. Specifically,
Fig.~\ref{f:lfs} shows that the LF slope seems to be distance-dependent
for the full extent of the cluster and for stars within the entire
observed brightness range.  Indeed, a functional relation between
$r_{\rm eff}$ and brightness seems to exist in Fig.~\ref{reffvsmag2} for
the entire cluster, with $r_{\rm eff}\simeq$~0.7\arcmin\ for $m_{\rm
F814W} \sim$~17~mag, but this trend is not supported statistically. The
small numbers of the brightest stars allow a solid statistical
interpretation of this plot only for stars of $m_{\rm
F814W}$~\gsim~21~mag. For these stars the LF slope becomes steeper
outwards. The advantage of our method lies in the fact that we are able
to define in a direct manner the distance of segregation for these
stars. This distance, derived from Fig.~\ref{reffvsmag2}, is $r_{\rm
eff}\simeq$~0.8\arcmin.

For Lindsay~91 a more obvious trend of $r_{\rm eff}$ as a function of
brightness can be seen for almost the entire observed magnitude range,
$m_{\rm F814W}$~\gsim~18.5~mag.  Also in this cluster, no clear
correlation between $r_{\rm eff}$ and $m_{\rm F814W}$ is observed for
the brighter stars due to their small numbers. The relation between
$r_{\rm eff}$ and $m_{\rm F814W}$ can be clarified easily, however,
for stars with $m_{\rm F814W}$~\gsim~21~mag. It is interesting to note
that from the comparison of the relation of $r_{\rm eff}$ vs. $m_{\rm
F814W}$ of Fig.~\ref{reffvsmag2} for Lindsay~91 with the corresponding
LFs for different distances from the cluster center, shown in
Fig.~\ref{f:lfs}, we see that both plots agree that significant
segregation is observed for stars with $m_{\rm
F814W}$~\gsim~21~mag. However, specific radial distances were selected
for the LFs of Fig.~\ref{f:lfs}, and therefore we cannot define an
accurate {\sl distance of segregation} for these stars, except that
segregation occurs at distances $r$~\gsim~40\arcsec. With our new
method we can accurately define the distance of segregation.  From
Fig.~\ref{reffvsmag2} we derive that this distance corresponds to the
effective radius of $r_{\rm eff} \sim$~0.72\arcmin~$\equiv$~43\arcsec,
which agrees very well with the results from the LFs method, but only
because of the arbitrary selection of this distance for this method.

It should be noted that the relation of $r_{\rm eff}$ vs. magnitude
derived for the MS stars (below the turn-off) is the most reliable
indicator of the degree of segregation of the studied clusters. In the
case of this relation for the brighter evolved stars (above the
turn-off) the situation is more complicated because the mass of these
stars changes very slowly with magnitude. Specifically, according to the
evolutionary models adopted here (Marigo et al. 2008) and the isochrone
of 6.3~Gyr, the difference in mass of stars evolving from the base of
the RGB to the AGB is much less than 0.01~M{\solar}, and thus the
corresponding values of $r_{\rm eff}$ refer to extremely narrow ranges
of mass per magnitude bin. As a consequence different magnitude bins
correspond to roughly the same mass and the relation of $r_{\rm eff}$
vs. magnitude for these stars provides indications of stratification of
of stars of different magnitude but nearly identical mass. Naturally,
this effect has consequences on the detection of {\sl mass segregation}.
In the case of the relation of $r_{\rm eff}$ vs. mass, the fact that the
evolved stars have small mass differences can improve the number
statistics, since a larger number of stars will correspond to the same
mass-bins. Indeed, considering that according to the isochrone model all
stars brighter than the turn-off have masses around 1~M{\solar}, if we
group together all these stars, then we should combine the first 10
magnitude bins shown in the plots of Fig.~\ref{reffvsmag2} into one. The
corresponding values of $r_{\rm eff}$ are 0.80\arcmin~$\pm$~0.06\arcmin\
for Kron~1 and 0.88\arcmin~$\pm$~0.07\arcmin\ for Lindsay~91; both
values are quite consistent with the derived trends.

In general, based on the above, with the method for detecting and
characterizing the effect of stellar stratification proposed here we can
distinguish between different degrees of segregation among star clusters
and compare our results from one cluster to another. The values of
$r_{\rm eff}$ derived by using only MS stars (below the turn-off) for
the studied clusters should be considered as the best tracers of mass
segregation.

\section{Conclusions}

We present a robust new method for the diagnosis of stellar
stratification in star clusters. This method uses the {\sl effective
radius}, which represents the observed Spitzer radius as estimated for
various stellar groups of different magnitude (or mass) ranges in the
same system. This radius is an important dynamical parameter, which
approximately describes the area of gravitational influence of a given
stellar group in the cluster. The proposed method detects stellar
stratification in star clusters: if a cluster is segregated, then the
effective radii for different stellar groups should be different from
one group to another. A relation between the effective radius of every
group and the corresponding selected magnitude (or mass) range can be
established to consistently characterize any stratification. As a result,
comparison among the results obtained for different clusters
can be made.

Stellar stratification in star clusters is usually quantified based on
either the surface density profiles of stars in different magnitude
ranges or the cluster LF (or MF) at different radial distances from
its center. Using the first method one may observe the magnitude
(mass) limit for the segregated stars and using the second the radial
distance where they are observed to be segregated. Both of these
results can be derived using the single application of the method
proposed here. Furthermore, to apply the `classical' methods for the
detection of mass segregation, certain complicated fitting procedures
must be applied, leading to model-dependent results. On the other
hand, our method is much more straightforward in its application, and
more precise in its results. This is mostly so because it is {\sl not}
necessary to (i) introduce any significant assumptions to estimate the
cluster parameters, and (ii) apply any fit to parameter correlations
and subsequently test the relation between the model output and the
observables. Both of these steps must be applied in the classical
approaches to characterize stellar stratification.

We present the application of this new diagnostic tool to two
mass-segregated clusters in the SMC, observed with {\sl HST}/ACS. The
application of the new method to space-based observations helped us to
establish a scheme for the comparison between the results for different
clusters and to define the important parameters required to characterize
stratification. These parameters are the {\sl magnitude} and {\sl radius
of segregation}, which specify the limits beyond which segregation
becomes stronger and more apparent. In addition, one may quantify the
differences between clusters through the {\sl slope of the relation
between the effective radius and the corresponding magnitude range},
which can be used as an indicator of the degree of segregation of each
cluster.

In each of the clusters observed with the {\sl HST}, considered here, it
was found that segregation is apparent for stars over the full observed
brightness range, but small-number statistics do not allow the detection
of a definite trend between $r_{\rm eff}$ and magnitude for several of
the brightest stellar groups. A comparable {\sl radius of segregation}
is observed for stars of $m_{\rm F814W}$~\gsim~21~mag in both clusters,
but the {\sl degree of segregation} seems to differ from one cluster to
the other, with Lindsay~91 exhibiting a steeper function of $r_{\rm
eff}(m_{\rm F814W})$ than Kron~1.

The continuous increase of $r_{\rm eff}$ as a function of brightness
away from the radius of segregation up to the observed limits of the
clusters shows that the fainter stars are distributed throughout a
larger volume, an observation reminiscent of the `evaporation' of star
clusters. Still, these results are based on only two clusters and can
therefore only be used rather qualitatively.

We conclude that the method we present here is simpler and more
accurate than previously developed methods. This is based, first and
foremost, on the more straightforward way in which the effective
radius is estimated and, secondly, on avoiding any model-dependent
fits. Instead, a direct correlation between the cluster's observables
is sufficient to exhibit any stellar stratification and to
parameterize its significance.

\acknowledgements

D. A. Gouliermis kindly acknowledges the support of the German Research
Foundation (Deu\-tsche For\-schungs\-ge\-mein\-schaft - DFG) through the
research grant GO~1659/1-1. We thank K. S. de~Boer for his contribution
to an earlier version of this paper, and R. Korakitis for his comments
regarding the projection of a spherical system. Based on observations
made with the NASA/ESA {\em Hubble Space Telescope}, obtained from the
data archive at the Space Telescope Science Institute. STScI is operated
by the Association of Universities for Research in Astronomy, Inc. under
NASA contract NAS 5-26555.




\begin{references}

\reference{} Bonnell, I. A., \& Davies, M. B. 1998, MNRAS 295, 691

\reference{} Bonatto, C., \& Bica, E.\ 2007, MNRAS, 377, 1301

\reference{} Brandl, B., Sams, B. J., Bertoldi, F., et al. 1996, ApJ, 466,
254

\reference{} Brocato, E., Di Carlo, E., \& Menna, G.\ 2001, A\&A, 374, 523

\reference{} Brocato, E., Castellani, V., Di Carlo, E., Raimondo, G., \&
Walker, A.~R.\ 2003, AJ, 125, 3111

\reference{} Chabrier, G., \& M\'era, D.\ 1997, A\&A, 328, 83

\reference{} D'Antona, F., \& Mazzitelli, I.\ 1983, A\&A, 127, 149

\reference{} de Grijs, R., Johnson, R.~A., Gilmore, G.~F., \& Frayn, C.~M.\
2002a, MNRAS, 331, 228

\reference{} de Grijs, R., Gilmore, G.~F., Johnson, R.~A., \& Mackey, A.~D.\
2002b, MNRAS, 331, 245

\reference{} de Grijs, R., Gilmore, G.~F., Mackey, A.~D., Wilkinson, M.~I.,
Beaulieu, S.~F., Johnson, R.~A., \& Santiago, B.~X.\ 2002c, MNRAS, 337,
597

\reference{} Elson, R. A. W., Fall, S. M., \& Freeman, K. C. 1987, ApJ, 323,
54

\reference{} Elson, R. A. W., Gilmore, G. F., Santiago, B. X., \& Casertano,
S.\ 1995, AJ, 110, 682

\reference{} Fischer, P., Pryor, C., Murray, S., Mateo, M., \& Richtler, T.\
1998, AJ, 115, 592

\reference{} Flower, P.~J., Geisler, D., Olszewski, E.~W., \& Hodge, P.\
1980, ApJ, 235, 769

\reference{} Dolphin, A.~E., et al. 2001, ApJ, 562, 303 

\reference{} Elson, R. A. W., Gilmore, G. F., \& Santiago, B. X. 1997, MNRAS,
289, 157

\reference{} Fischer, P., Pryor, C., Murray, S., Mateo, M., \& Richtler, T.\
1998, AJ, 115, 592

\reference{} Gouliermis, D., Kontizas, M., Korakitis, R., et al. 2000, AJ,
119, 1737

\reference{} Gouliermis, D., Keller, S.~C., Kontizas, M., Kontizas, E., \&
Bellas-Velidis, I.\ 2004, A\&A, 416, 137

\reference{} Hillenbrand, L. A., \& Hartmann, L. W. 1998, ApJ, 492, 540

\reference{} Holtzman, J. A., et al. 1999, AJ, 118, 2262

\reference{} Johnson, J.~A., Bolte, M., Stetson, P.~B., Hesser, J.~E., \&
Somerville, R.~S.\ 1999, ApJ, 527, 199

\reference{} Keller, S.~C., Bessell, M.~S., \& Da Costa, G.~S.\ 2000, AJ,
119, 1748

\reference{} King, I. R. 1965, ApJ, 142, 387

\reference{} King, I. R. 1966, AJ, 71, 64

\reference{} King, I.~R., Sosin, C., \& Cool, A.~M.\ 1995, ApJL, 452, L33

\reference{} Keller, S.~C., Bessell, M.~S., \& Da Costa, G.~S.\ 2000, AJ,
119, 1748

\reference{} Kontizas, M., Hatzidimitriou, D., Bellas-Velidis, I.,
Gouliermis, D., Kontizas, E., \& Cannon, R.~D.\ 1998, A\&A, 336, 503

\reference{} Kron, G. E.\ 1956, PASP, 68, 125 

\reference{} Kroupa, P.\ 1998, MNRAS, 298, 231

\reference{} Kroupa, P., \& Tout, C. A.\ 1997, MNRAS, 287, 402

\reference{} Kroupa, P., Tout, C. A., \& Gilmore, G. F.\ 1990, MNRAS, 244,
76

\reference{} Kroupa, P., Tout, C. A., \& Gilmore, G. F.\ 1993, MNRAS, 262,
545

\reference{} Lightman, A. P. \& Shapiro, S.L., 1978, Rev. Mod. Phys. 50, 437

\reference{} Lindsay, E. M.\ 1958, MNRAS, 118, 172 

\reference{} Mackey, A.~D., \& Gilmore, G.~F.\ 2003a, MNRAS, 338, 120 

\reference{} Mackey, A.~D., \& Gilmore, G.~F.\ 2003b, MNRAS, 340, 175 

\reference{} Mackey, A.~D., \& Gilmore, G.~F.\ 2004, MNRAS, 352, 153

\reference{} Mackey, A.~D., Payne, M.~J., \& Gilmore, G.~F.\ 2006, MNRAS,
369, 921

\reference{} Ma{\'{\i}}z Apell{\'a}niz, J., \& {\'U}beda, L.\ 2005, ApJ,
629, 873

\reference{} Marigo, P., Girardi, L., Bressan, A., Groenewegen, M.~A.~T.,
Silva, L., \& Granato, G.~L.\ 2008, A\&A, 482, 883

\reference{} Meylan, G. \& Heggie, D.C. 1997, A\&A Rev. 8, 1

\reference{} Murray, S. D., \& Lin, D. N. C. 1996, ApJ 467, 728

\reference{} Pandey, A.~K., Mahra, H.~S., \& Sagar, R.\ 1992, Bulletin of the
Astronomical Society of India, 20, 287

\reference{} Plummer, H. C. 1911, MNRAS, 71, 460

\reference{} Rochau, B., Gouliermis, D.~A., Brandner, W., Dolphin, A.~E., \&
Henning, T.\ 2007, ApJ, 664, 322

\reference{} Sagar, R., Myakutin, V.~I., Piskunov, A.~E., \& Dluzhnevskaya,
O.~B.\ 1988, Bulletin of the Astronomical Society of India, 16, 87

\reference{} Scaria, K.~K., \& Bappu, M.~K.~V.\ 1981, Journal of Astrophysics
and Astronomy, 2, 215 (Corrigendum: 1981, Journal of Astrophysics and
Astronomy, 2, 439)

\reference{} Santiago, B., Beaulieu, S., Johnson, R., \& Gilmore, G.~F.\
2001, A\&A, 369, 74

\reference{} Sirianni, M., Nota, A., De Marchi, G., Leitherer, C., \&
Clampin, M.\ 2002, ApJ, 579, 275

\reference{} Smecker-Hane, T. A., Cole, A. A., Gallagher, J. S., \& Stetson,
P. B. 2002, ApJ, 566, 239

\reference{} Spitzer, L. Jr. 1958, ApJ, 127, 17 

\reference{} Spitzer, L. Jr. 1969, ApJ, 158, L139 

\reference{} Stanghellini, L., Villaver, E., Shaw, R.~A., \& Mutchler, M.\
2003, ApJ, 598, 1000

\reference{} Subramaniam, A., Sagar, R., \& Bhatt, H. C. 1993, A\&A, 273, 100

\reference{} Testa, V., Ferraro, F.~R., Chieffi, A., Straniero, O., Limongi,
M., \& Fusi Pecci, F.\ 1999, AJ, 118, 2839

\reference{} Xin, Y., Deng, L., de~Grijs, R., Mackey, A.~D., \& Han Z. 2008,
MNRAS, 384, 410


\end{references}
\end{document}